\documentclass[pra,twocolumn,superscriptaddress,amsmath,amssymb,floatfix]{revtex4-2} 
\usepackage{soul} % For highlighting
\usepackage{graphicx} % Include figure files
\usepackage{dcolumn}  % Align table columns on decimal point
\usepackage{bm}       % bold math
\usepackage[colorlinks=true,linkcolor=blue,urlcolor=blue,citecolor=blue,linkcolor=blue,pdfstartview=FitH]{hyperref}

\makeatletter
\newcommand*{\rom}[1]{\expandafter\@Slowromancap\romannumeral #1@}
\makeatother
\usepackage{enumerate}
\usepackage{amsmath}
\usepackage{color}
\usepackage{cases}
\usepackage{microtype}
\usepackage{siunitx}
\usepackage{subdepth}
\usepackage[utf8]{inputenc}
\usepackage[T1]{fontenc}
\usepackage{txfonts} %PRA-like font (math)
\begin{document}

\title{Machine learning classification of two-dimensional vortex configurations}
\author{Rama Sharma}
\affiliation{Optical Sciences Centre, Swinburne University of Technology, Melbourne 3122, Australia}
\author{Tapio P. Simula}
\affiliation{Optical Sciences Centre, Swinburne University of Technology, Melbourne 3122, Australia}

\begin{abstract}
We consider computer generated configurations of quantised vortices in planar superfluid Bose--Einstein condensates. We show that unsupervised machine learning technology can successfully be used for classifying such vortex configurations to identify prominent vortex phases of matter. The machine learning approach could thus be applied for automatically classifying large data sets of vortex configurations obtainable by experiments on two-dimensional quantum turbulence.
\end{abstract}

\maketitle

%%%%%%%%%%%%%%%%%%%%%%%%%%%%%%%%%%%%%%%%%%%%%%%%%%%
\section{Introduction}

Statistical mechanics is one of the cornerstones of modern physics \cite{gibbs2014elementary, gogolin2016equilibration, sethna2021statistical}. The key underlying principle is statistical equivalence of the myriad of microstates that all share common thermodynamic properties such as configuration energy or temperature, and collectively represent the macroscopically observable phenomena. Typically, the macroscopic state variables are given as input parameters and the corresponding microstates are obtained as solutions to the underlying model Hamiltonian. In this work, we are interested in a reverse process where the microstates are given as measurement outcomes which we wish to categorize into distinct macrostates based on their statistical similarity, without any knowledge of the underlying Hamiltonian.

The Onsager model of two-dimensional turbulence \cite{Onsager1949statistical} is a statistical mechanics description of point-like vortices where each vortex configuration corresponds to a particular statistical microstate of the fluid. The model is particularly well suited for modeling vortices and their statistical behaviour in superfluids where the vorticity of the fluid is quantised. The applicability of the Onsager model has been verified by recent experiments on two-dimensional quantum turbulence (2DQT) in superfluid Bose--Einstein condensates \cite{gauthier2019giant,johnstone2019evolution}. 

In typical cold atoms experiments the quantitative information of the physical system is extracted from images of the atom density distribution \cite{Varenna1999a}. Each such experimental image corresponds to a single representative microstate of the system's thermodynamic state and if quantised vortices are present in the system, their positions can be read off from such images. This raises the question whether it would be possible to categorize such experimental data using machine learning protocols into ensembles of statistically equivalent microstates, for instance in order to detect distinct phases of matter supported by the system?

The remarkable successes of artificial neural networks when applied to the problems of image recognition, image classification and natural language processing has prompted interdisciplinary efforts to investigate how a broader range of scientific problems might benefit from deploying these new tools. This has led to implementations of machine learning methods, for instance, to identify symmetry-broken phases in the field of classical statistical physics \cite{carrasquilla2017machine,van2017learning, wetzel2017machine}, and in some cases neural networks have even been shown to be able to learn an order parameter or other thermodynamical parameters \cite{carrasquilla2017machine,wetzel2017machine}. More recently, the machine learning methodology has found applications in the realm of physics problems such as identifying phase transitions of many-body systems \cite{miyajima2021machine,dong2019machine, carleo2019machine, shirinyan2019self, kaming2021unsupervised,zhang2018machine,zhang2019machine,shiina2020machine,torlai2016learning,carleo2017solving,wang2016discovering}, topological systems \cite{richter2018machine,beach2018machine,zhang2017quantum, huembeli2018identifying,rodriguez2019identifying}, and  finding quantum enhanced learning algorithms \cite{dunjko2018machine,biamonte2017quantum,schuld2015introduction}. 

Appropriately designed and trained supervised deep learning procedure has been applied to spin and vortex configurations in the two-dimensional XY model \cite{beach2018machine} to identify the Kosterlitz--Thouless (KT) transition \cite{kosterlitz1973ordering}. However, labeled training sets, which are mandatory for supervised learning, are not always easily attainable. For this reason, the unsupervised machine learning methods, that do not rely on prior knowledge, may offer significant benefits over supervised learning methods \cite{greplova2020unsupervised, fukushima2019featuring}. Inspired by the application of the supervised learning to classify the KT transition \cite{beach2018machine,miyajima2021machine}, it is natural to ask whether unsupervised neural networks would be capable of identifying a variety of vortex phases of matter and quantum turbulent flow states in two-dimensional Bose--Einstein condensates? Few unsupervised learning techniques have been previously applied to the XY model. The principal component analysis (PCA) method \cite{jolliffe2002principal} has been performed on spin configurations \cite{wetzel2017unsupervised,hu2017discovering, wang2017machine} but even when learning with the vorticity field directly, the PCA was found to be unable to identify the transition point corresponds to the vortex--antivortex unbinding \cite{hu2017discovering}. Additionally, most of these previous approaches need the prior information before processing for example; the previously known number of phases and approximate transition temperature value \cite{richter2018machine,zhang2019machine}. Hence such aspects can be complicated to implement the generalised idea via the machine learning technique to detect these phase transitions. In contrast, our work presents machine learning approach which employs the bag of feature function for feature extraction and unsupervised self-organising map (SOM) algorithm for classification and is able to detect the ordered and disordered sides of the vortex binding-unbinding transition without prior labeling.

In this paper, we apply unsupervised machine learning strategy to the task of identifying vortex phases of matter of the two-dimensional Onsager point vortex model. Our main goal is to test whether it is possible for an artificial neural network, trained only on the features extracted from the vortex configurations, to learn distinct vortex phases of matter that are thermodynamically defined in terms of external macrostate properties such as temperature. The rest of this paper is organised as follows. The Sec.~II provides the theoretical background on the Onsager's point vortex model of two-dimensional turbulence and outlines the numerical methods employed, such as the Monte Carlo method and machine learning model. The Sec.~III begins by providing benchmark results using unsupervised machine learning approach showing successful classification of states of a two same sign vortex system. Specifically, we show that the unsupervised learning is able to identify the analytically predicted topological phase boundary of this system \cite{Navarro2013a,murray2016hamiltonian}. We then consider experimentally relevant point vortex configurations of twenty polarised (single sign circulation for all vortices) and forty neutral (equal number of vortices and antivortices) vortex systems and show that the unsupervised machine learning is able to successfully identify both the positive vortex temperature Kosterlitz--Thouless transition as well as the negative vortex temperature Onsager vortex condensation transition. These findings show promise for applying machine learning models for exploring experimental data sets involving vortex configurations. We close the paper in Sec.~IV by summarising our findings with concluding remarks.

%%%%%%%%%%%%%%%%%%%%%%%%%%%%%%%%%%%%%%%%%%%
\section{Theory}
%%%%%%%%%%%%%%%%%%%%%%%%%%%%%%%%%%%%
Consider a Navier--Stokes equation, %\cite{temam2001navier},
\begin{equation}
\frac{\partial {\mathbf{v}}}{\partial t} + (\mathbf{v}\cdot\nabla)
\mathbf{v} = -\frac{\nabla {p}
}{\rho} + \nu\nabla^2\mathbf{v},
\label{eq:1}
\end{equation}
which describes the flow of a Newtonian incompressible fluid that satisfies the continuity equation $\partial_t \rho+\nabla\cdot(\rho\mathbf{v})=0$. In Eq.~(\ref{eq:1}) $\mathbf{v}(\mathbf{r}, t)$ is the fluid velocity field, $\rho$ is the fluid density, $p$ is the fluid pressure, $\nabla^2$ is the Laplacian operator and $\nu$ is the kinematic viscosity. In rare cases, such as one-dimensional flow and creeping flow, the Navier--Stokes equation for $\mathbf{v}(\mathbf{r}, t)$ can be solved analytically. However, the nonlinearity in turbulent fluids arising due to the convective acceleration of the fluid makes analytical solutions impossible in general. Moreover, an accurate numerical solution of Eq.~(\ref{eq:1}) is difficult to achieve due to the vast range of length scales and number of degrees of freedom involved \cite{cant2001sb}. Fortuitously, a dramatic simplification can be achieved by considering the fluid's vorticity field, $\boldsymbol{\omega}(\mathbf{r}, t)=\nabla\times\mathbf{v}(\mathbf{r}, t)$, instead of the velocity field. In particular in the case of planar superfluids, the former can in many cases be well approximated by only a small number of point vortices. 

 %%%%%%%%%%%%%%%%%%%%%%%%%%%%%%%%%%%%%%%%%%%
\subsection{Point vortex dynamics}
One of the conceptual benefits of two-dimensional fluid dynamics comes from the fact that the flow field is confined to a plane. Consequently, the vortices cannot bend or expand the way they can in three dimensional flows. The vortices can be modeled as point-sources of rotating fluid flow in the limit where the vortices are well separated. In such a case, the vorticity $\mathbf{\omega}(\mathbf{r})$, which is normally smoothly distributed over the fluid, may be obtained by summing over the vortices according to
\begin{equation}
\boldsymbol{\omega}(\mathbf{r}) =\nabla \times \mathbf{v}= \sum_{i}^{N_v} \Gamma{_i} \hat{\mathbf{e}}_z\delta (\mathbf{r}-\mathbf{r}_i),
 \label{eq:curlv}
\end{equation}
where the point vortices with circulations $\Gamma_i$ are located at positions $\mathbf{r}_i$ for $i\in \{1,2,\ldots,N_v\}$ with $N_v$ the total number of vortices. The velocity field of a fluid flow around a single vortex at location $r$ is
\begin{equation}
\mathbf{v}_i(\mathbf{r}) = \frac{\Gamma{_i}}{2\pi}\frac{1}{|\mathbf{r}-\mathbf{r}_i|}\hat{\theta}_i,
 \label{eq:3}
\end{equation}
where the azimuthal co-ordinate axes $\hat{\theta}_i$  are centred on the vortex cores and the total fluid velocity field
\begin{equation}
\mathbf{v}(\mathbf{r})=\sum_{i}^{N_v} \mathbf{v}_i(\mathbf{r})
 \label{eq:4}
\end{equation}
due to many vortices is a simple superposition of the individual vortex velocity fields.
The point vortex approximation significantly simplifies the modeling of a continuous 2D fluid, as the velocity of the fluid at each point in space can be mapped by the net flow of all vortices within the fluid. The vortices are positioned in such a way that the superposition of their regular circulating flows better mimics the fluid's complete velocity field \cite{aref2007point}. This description has the advantage of allowing the continuous fluid to be replaced by $N_v$ zero spatial extent points (vortices) of well-defined locations $(x_i, y_i)$ and circulations $\Gamma_i$, which carry the full fluid flow information. The dynamics of the vortices maps onto the dynamics of the fluid particles and each vortex moves with the fluid velocity induced by all other vortices within the fluid at its location. Hence, each vortex’s motion is measured by the relative location and strength of all other vortices \cite{weissmann2014hamiltonian}.

The point vortex approximation is particularly well suited for modeling the dynamics of vortices in superfluid Bose--Einstein condensates (BECs) for which Eq.~(\ref{eq:curlv}) is accurately satisfied. This leads to the point vortex model (PVM) for the dynamics of the vortices, which utilises a set of coupled ordinary differential equations to describe the vortex dynamics by considering the interactions among vortices and related boundary conditions in trapping potential. For most of this work we make the use of a \emph{uniform trap} potential. We provide brief description of uniform trapping and the dynamics of two and three vortices in position space in the following.

\subsubsection{Vortex equations of motion}
The so-called box potentials have become commonly utilised in cold atom experiments \cite{chomaz2015emergence, tempone2017high, johnstone2019evolution,gauthier2019giant}. Considering vortices in a unit disk geometry, each vortex is accompanied by a single image vortex of charge $\bar s_i = - s_i$ positioned outside the fluid boundary at a location $ \bar{\mathbf{r}}_i = \mathbf{r}_i R^2 / |\mathbf{r}_i|^2$ \cite{pointin1976statistical}. 
The equations of motion for such point  vortices are \cite{Onsager1949statistical},
 \begin{align}
 \label{eq:pvm}
hs_i \frac{\partial x_i}{\partial t} =  \frac{\partial H}{\partial y_i} \hspace{0.4cm} \textrm{and} \hspace{0.4cm}    hs_i\frac{\partial y_i}{\partial t} =  -\frac{\partial H}{\partial x_i},  
\end{align}
where the energy of the vortex configuration is

\begin{align}
\label{eq:H}
H &= \alpha k_B\sum_{i=1}^{N_v} s_i^2\ln{(1-r_i^2)} - \alpha k_B\sum_{j<i}^{N_v} s_is_j\ln{(r_{ij}^2)} \\ 
&+ \alpha k_B\sum_{j<i}^{N_v} s_is_j\ln{(1-2x_ix_j- 2y_iy_j + r_i^2r_j^2)}. \notag
\end{align}
In Eq.~(\ref{eq:H}), $\alpha = \rho_s \kappa^2/4\pi k_B$ where $\rho_s$ is the superfluid density and $\kappa = h/m$ is the quantum circulation with $h$, $m$ and $k_B$ denoting the Planck’s constant, the atom mass and Boltzmann constant, respectively. The $r_i^2 = x_i^2 + y_i^2$ with $x_i = {\rm Re}({\bf z}_i)$ and $y_i = {\rm Im}({\bf z}_i)$ are the Cartesian coordinates of $i$th vortex and are expressed in terms of complex numbers ${\bf z}_i$ in a system of dimensionless radius $R=1$ with circulation winding number $s_i = \pm 1$. The first term of Eq.~(\ref{eq:H}), corresponds to the interaction of the vortices with their images. The second term describes the logarithmic long-range two-dimensional Coulomb interaction between the vortices, and last term is the interaction of the real vortices with the images of all other vortices in the system.

 %%%%%%%%%%%%%%%%%%%%%%%%%%%%%%
\begin{figure}[t]
\centering
  \includegraphics[width=\columnwidth]{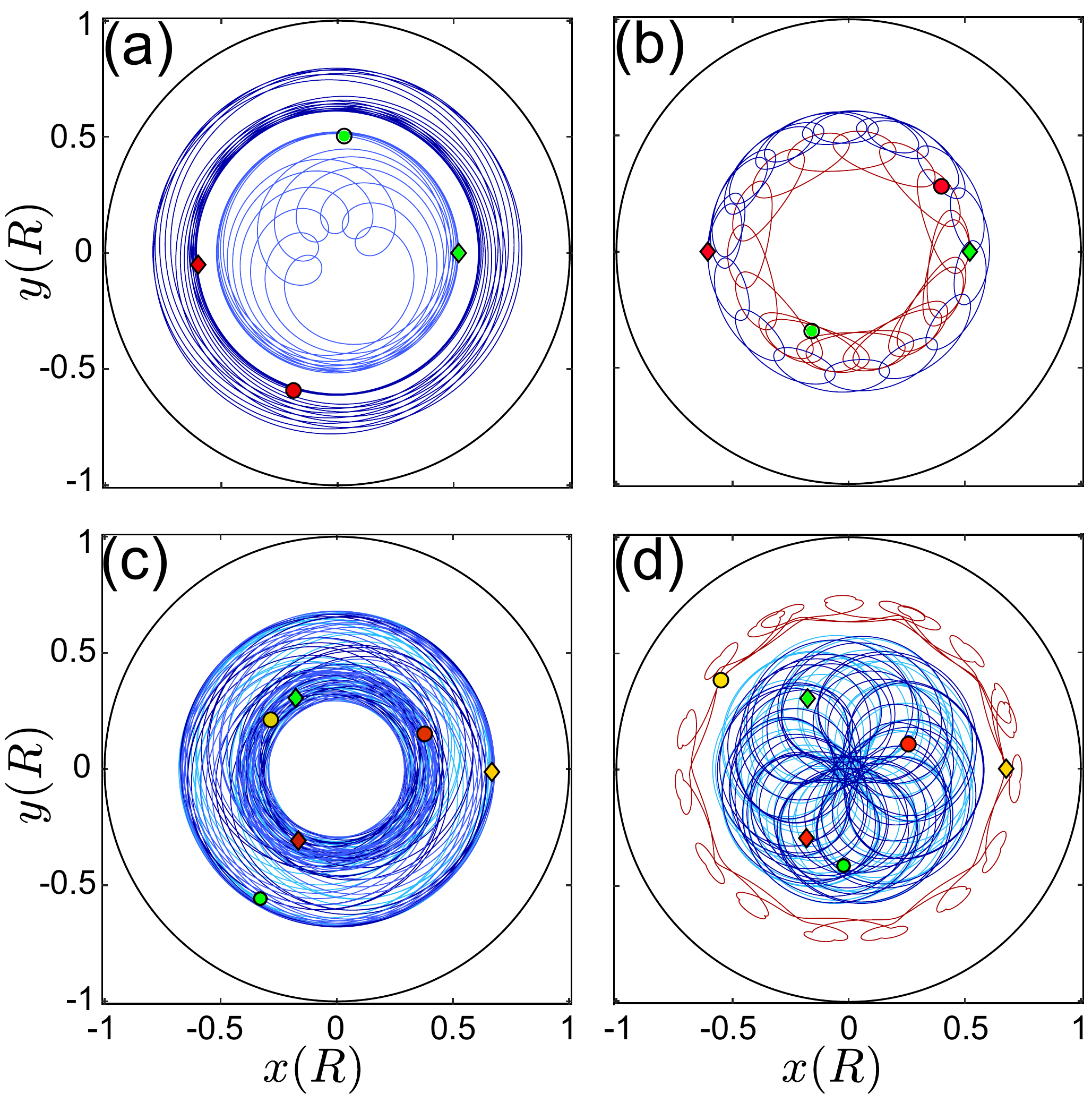}
  \caption{Representative dynamics of two and three vortices in a uniform trap. Frames (a) and (b) represent the asymmetric rigidly-rotating states for the given initial parameters $\phi/\pi = 0.274, L = 0.64R^2$ and $ \theta_{21} = \pi$ for two same sign $s_1=1=s_2$ and opposite sign $s_1=1=-s_2$ vortices, respectively. Frame (c) shows chaotic dynamics of three same sign $s_1=s_2=s_3=1$ vortices for the initial parameters $\phi/\pi = 0.45, L = 0.70R^2$ and $ \theta/\pi = 0.204$. Frame (d) represents the dynamics for the case $s_1=s_2=1=-s_3$ for the same initial parameters as in frame (c). In all frames, the initial and final positions of the vortices are denoted by diamonds and circular markers, respectively.}
\label{Fig:F1}
\end{figure}
 
Substituting Eq.~(\ref{eq:H}) into  Eq.~(\ref{eq:pvm}) yields the point vortex equations of motion 
\begin{equation}
\label{eq:pvmuni}
\mathbf{u}_i =  \frac{\hbar}{m}\left (\sum_{j\neq
 i}^{N_v} s_j\mathbf{\hat{e}}_z\times{\frac{\mathbf{r}_i-\mathbf{r}_j}{|\mathbf{r}_i-\mathbf{r}_j|^2}} + \sum_{j=1}^{N_v} \bar s_j\mathbf{\hat{e}}_z\times{\frac{\mathbf{r}_i-\bar{\mathbf{r}}_j}{|\mathbf{r}_i-\bar{\mathbf{r}}_j|^2}}\right), 
 \end{equation}
 that model the vortex dynamics in a hard walled uniform disk trapped Bose-Einstein condensate \cite{groszek2018motion}, with ${\bf u}_i$ the velocity of the $i$th vortex.
 
 A similar PVM was obtained phenomenologically by considering an oblate harmonically trapped Bose--Einstein condensate where the motion of the vortices is restricted within the Thomas-Fermi radius, $R > |z_i|$ of the BEC \cite{middelkamp2011guiding, moon2015thermal, Navarro2013a}. The resulting equations of motion 
\begin{equation}
-i \mathbf{\Dot{z}}_i =  R^2\Omega_0\frac{s_i\mathbf{z}_i} {R^2- |\mathbf{z}_i|^2} + R^2 \Omega_{\rm int}\sum_{j\neq
 i}^N s_j \frac{\mathbf{z}_i- \mathbf{z}_j}{|\mathbf{z}_i- \mathbf{z}_j|^2} \label{eq:pvmsho}
 \end{equation}
have two further phenomenological constants $\Omega_0$ and $\Omega_{\rm int}$, where the former is the orbital angular frequency of a vortex which orbits around the trap centre \cite{Navarro2013a,fetter2001vortices,freilich2010real}, and the latter is the angular frequency that determines the inter vortex interaction strength. 

%%%%%%%%%%%%%%%%%%%%%
\subsubsection{Position-space dynamics of two vortices}

To demonstrate the basics of vortex dynamics we first consider a system of two vortices, $N_v = 2$, of unit circulation in a uniform trap. The length and time are measured in the units of the system radius $R$ and inverse angular frequency $\Omega_0^{-1}$, respectively. All possible two-vortex configurations in this system can be characterised by three observables; the vortex angular momentum $L = r_1^2 + r_2^2$, not to be confused with the angular momentum of the fluid, the angle $\phi = \tan^{-1}(r_2/r_1)$, and the azimuthal angle $\theta_{21} = \theta_2 - \theta_1$, subtended by the position vectors of the two vortices. 

An experimental and theoretical investigation of two-vortex systems in a harmonic trap with $\theta_{21} = \pi$ and $\Omega_{int}/\Omega_0 = 0.1$ was conducted in Ref.~\cite{Navarro2013a, murray2016hamiltonian}. Here we chose all parameters according to \cite{Navarro2013a, murray2016hamiltonian}, except for setting $\Omega_{int}/\Omega_0 = 1$ for uniform trap.
%%%%%%%%%%%%%%%%%
We first place the two vortices on opposite sides of the trap center with $\phi/\pi = 0.274, L = 0.64R^2$ and $ \theta_{21} = \pi$ and position coordinates $(x_i,y_i)\in\mathbb{R},r_i< R$. The subsequent evolution of the vortex configurations are illustrated in figure~\ref{Fig:F1}(a) for vortices of the same sign, $s_1=1=s_2$, and in figure~\ref{Fig:F1}(b) for the case of vortices of opposite sign $s_1=1=-s_2$.
Frame (a) shows that for this initial configuration the orbits traced out by the two same sign vortices are confined to the separate phase space regions and their orbits never intersect. Frame (b) exhibits the two opposite sign vortices dynamics for the same initial configuration as in (a). The paths of the same sign vortices are shown using shades of blue in frame (a) and the opposite sign vortex path is shown using red color in frame (b). In both frames the initial and final positions of vortices are indicated by diamonds and circular markers, respectively. The vortex paths shown in Fig.~\ref{Fig:F1} are tracked for the duration of $60\; \Omega_0^{-1}$.
 %%%%%%%%%%%%%%%%%%%

\subsubsection{Position-space dynamics of three vortices}
By introducing the third vortex in the system, an additional pair of conjugate variable degrees of freedom is added to the vortex phase space. Consequently, the system is no longer integrable in general. Hence it is to be expected that for some initial conditions the three vortex system results in chaotic dynamics. To illustrate the three vortex case we consider a system with $|s_i|=1$ and phase space spanned by three coordinates $\theta = \cos^{-1}(r_3/\sqrt{L})$, $\phi = \tan^{-1}(r_2/r_1)$ and $ L = r_1^2 + r_2^2 + r_3^2$ with the constraint of $r_1 = r_2$ such that $\phi = \pi/4$ to constrain the available degrees of freedom. Changing $\theta$ corresponds to changing the ratio of the third vortex radius $r_3$ to the the radii of the remaining two vortices. The relative vortex locations were chosen in such a way that the angular separation of vortices $i$ and $j$ is $\alpha_{ij} = 2\pi/3$. These constraints ensure that for $r_1 = r_2 = r_3$, a fully symmetric rigidly rotating state will be obtained. The dynamics of three same sign vortices are shown in Fig.~\ref{Fig:F1}(c), where the initial and final positions of the third vortex are shown in yellow  diamond and circular markers, respectively. The dynamics  for the case in which one of the three vortices has an opposite sign of circulation compared to the other two is shown in Fig.~\ref{Fig:F1}(d). 

 To compute the point vortex dynamics in uniform and harmonic traps, Eqns~\eqref{eq:pvmuni} and \eqref{eq:pvmsho} were integrated using \texttt{MATLAB} function \texttt{ode113} with relative error tolerance and absolute error tolerance both set to $10^{-13}$. Each simulation was initialized by setting the initial positions and circulations of the vortices. The accuracy of the integration was confirmed by monitoring the values of the conserved quantities $H$ and $L$.

\begin{figure}[!t]
    \centering
    \includegraphics[width=\columnwidth]{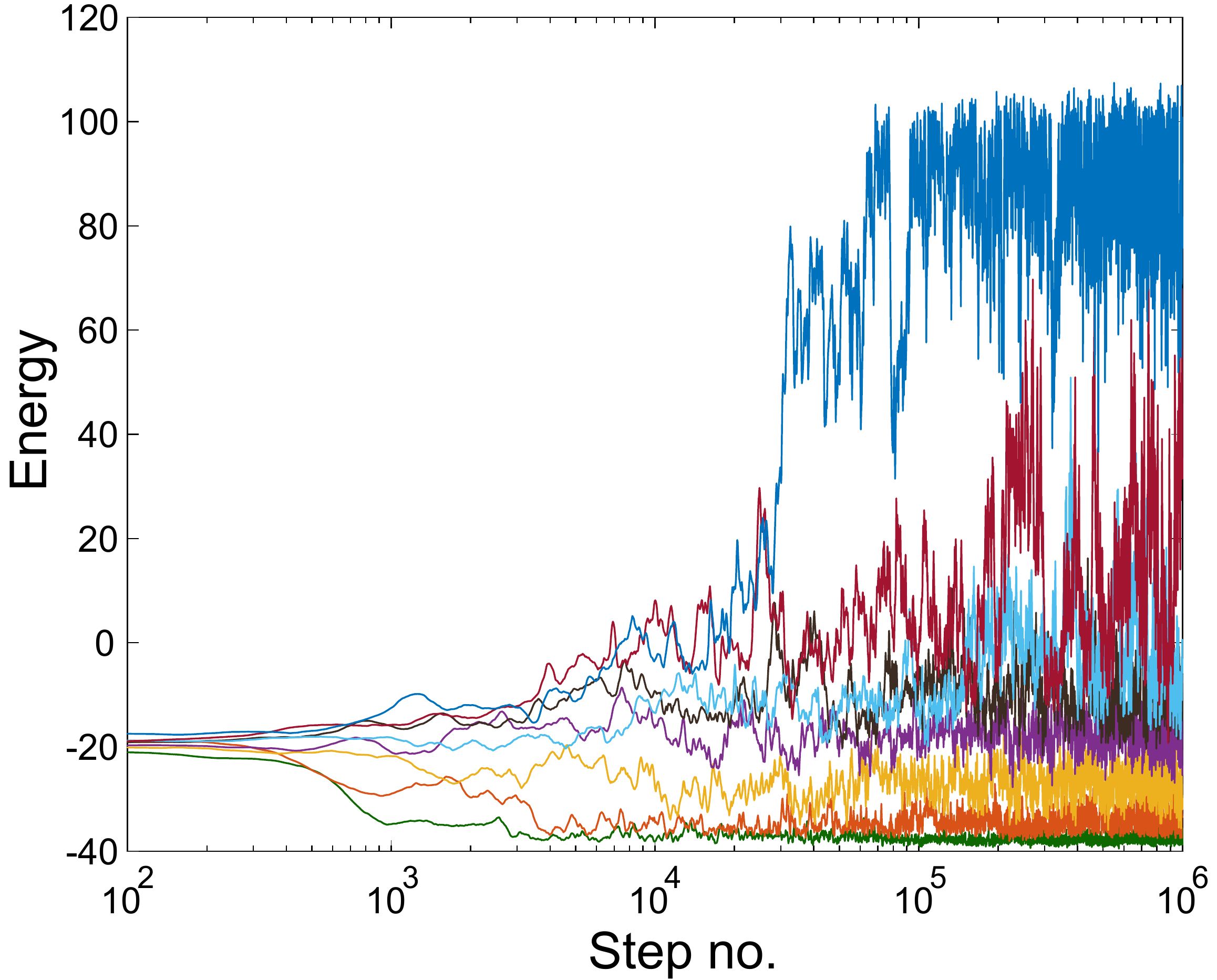}
\caption{Energy per vortex of 10 vortices as functions of the Monte Carlo step number for $8$ temperature points in the range $\beta = [1.4 \beta_{\rm{BKT}}, \;-1.4 \beta_{\rm{EBC}} ]$  (green to blue) in steps of $-0.4$. 
\label{Fig:F2}}
\end{figure}

\subsection{Monte Carlo thermodynamics}

Since we are particularly interested in the statistical properties of two-dimensional vortex configurations, we have implemented a Monte Carlo sampling method following Refs.~\cite{groszek2018vortex, valani2018einstein}. Briefly, the algorithm takes the initial positions and circulations of $N_v$ vortices and a vortex temperature $T_v$ as input parameters and returns a set of statistically equivalent equilibrium vortex configurations.

To initiate the algorithm we generate random initial positions for the fixed number of $N_v$ vortices. In each step of the algorithm an attempted move of one randomly selected vortex is made and the move is accepted or rejected probabilistically based on a temperature dependent weight function $\eta$. In this work a Boltzmann factor $\eta = \exp(-H/k_{B}T_v)$, where $k_{B}$ is the Boltzmann constant and $T_v$ is the vortex temperature, is used. The inverse temperature $\beta = 1/(k_{\rm B}T_v)$ characterises the equilibrium vortex configurations. The energy $H$ is determined from the point-vortex Hamiltonian Eq.~(\ref{eq:H}) of a uniform fluid inside a circular domain of radius $R$. The core radius of the vortices is set to be $0.008\;R$. In order to account for the effect of finite core size of real superfluid vortices, a constraint was placed when generating the vortex configurations to avoid vortices from falling too close to each other or to the system boundary. A minimum intervortex separation of twice the vortex core radius, and a minimum vortex-boundary separation of one vortex core radius was implemented. To ensure fair sampling of the vortex configurations, a variety of observables were monitored. Figure~\ref{Fig:F2} shows the energy of the vortex configuration as functions of Monte Carlo step number for 8 different temperatures. All of the 8 runs were initiated with the same initial configuration that was prepared by drawing the vortex positions randomly from a uniform distribution. These systems were deemed to have reached equilibrium after $(10^5)$ steps. All Monte Carlo calculations were therefore first run for an initial burn in of $10^5$ steps at each temperature. Subsequently, at every temperature, $1000$ vortex configurations were sampled uniformly from the total of $10^6$ microstates generated.

%%%%%%%%%%%%%%%%%%
\subsubsection{Clusters, dipoles and free vortices}
\begin{figure}
   \centering
    \includegraphics[width=\columnwidth]{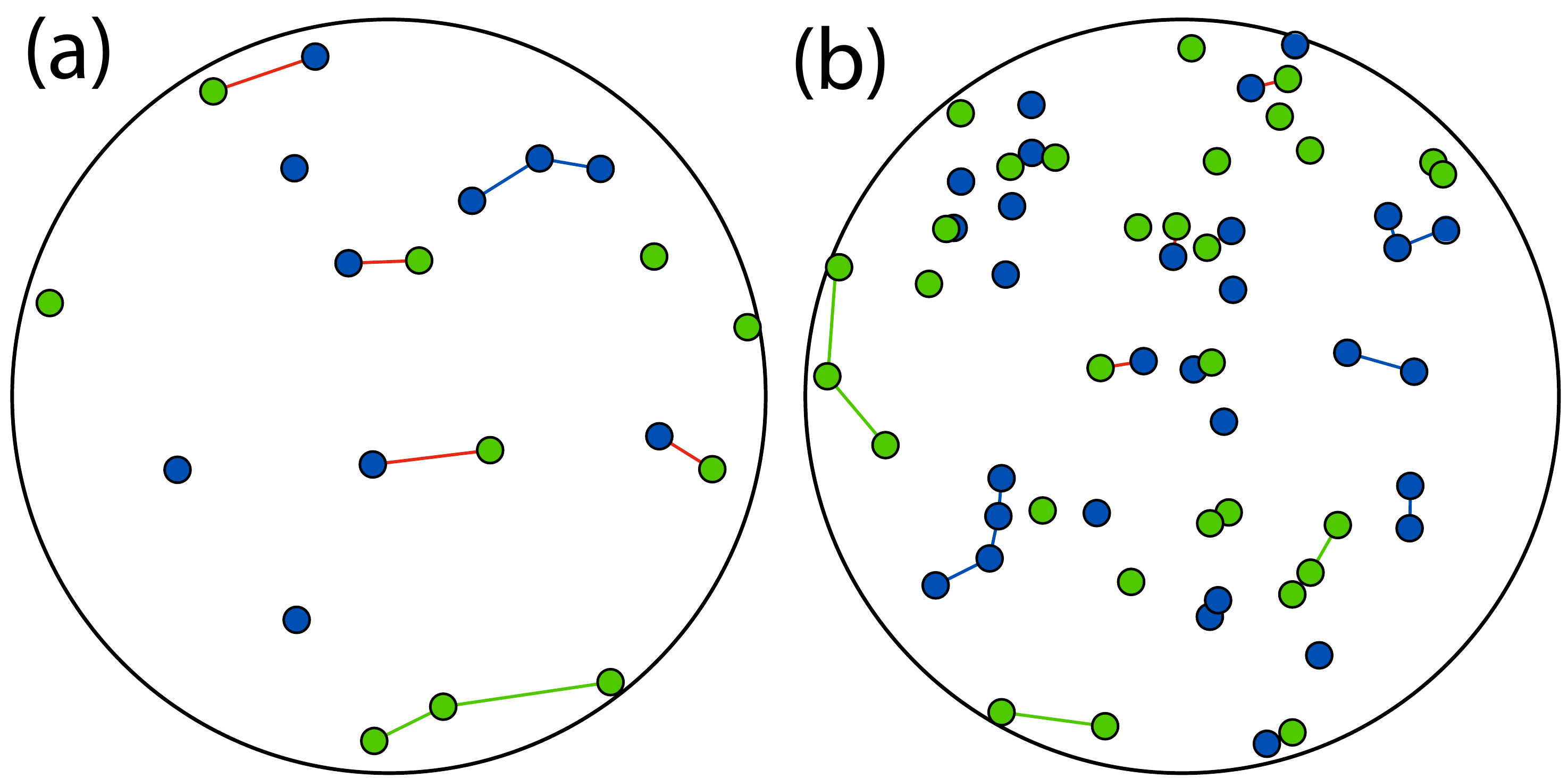}
    \caption{Classified vortex configurations for two infinite vortex temperature systems. (a) and (b) illustrate the neutral vortex configurations of 20 and 60 vortices, respectively. Blue and green  circular markers in both panels indicate the vortices and antivortices, respectively. The red line joins the dipole pairs. Blue and green lines join the clusters of vortices and antivortices, respectively. The vortices which are not joined by any of these lines are classified as free vortices.} 
    \label{Fig:F3}
\end{figure}

To quantitatively detect different vortex configurations we used a vortex classification algorithm \cite{reeves2013inverse,valani2018einstein}. The algorithm classifies vortices based on their spatial configuration by measuring all intervortex distances and assigning the vortices into three categories: same-sign clustered vortices, vortex dipoles, and free vortices. Figure~\ref{Fig:F3} shows the outcome of this classification algorithm for exemplary configurations of 20 and 60 vortices generated by the Monte Carlo algorithm at infinite $(\beta = 0)$ vortex temperature. 

%%%%%%%%%
\subsection{Vortex fluid phase transitions}
\begin{figure}
   \centering
    \includegraphics[width=\columnwidth]{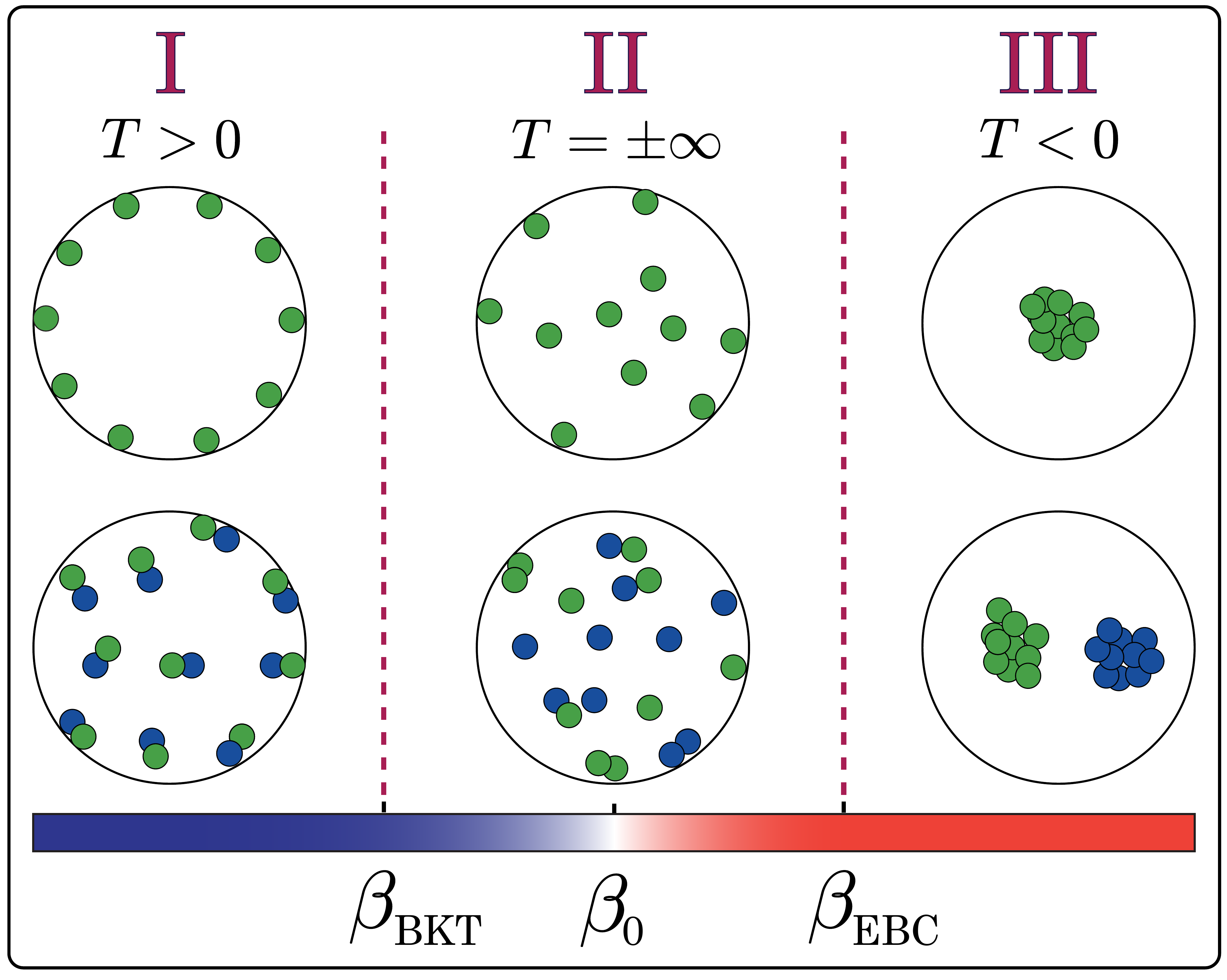}
    \caption{Illustration of the three point vortex phases of matter as functions of inverse temperature $\beta$. The top row shows configurations of 10 vortices all having the same sign of circulation. 
    The bottom row shows configurations of 20 vortices with equal numbers of clockwise and anti-clockwise circulations. The three phases shown are the low entropy positive temperature phase I, high entropy disordered phase II, and low entropy negative temperature phase III. The two extremes are separated from the disordered phase by the critical temperatures $\beta_{\rm BKT}$ and $\beta_{\rm EBC}$, which are marked by dashed vertical lines.}
    \label{Fig:F4}
\end{figure}

The point vortex statistical thermodynamics sets a framework for understanding the vortex fluid behaviour within the point vortex model. The model has two prominent transition temperatures corresponding to the positive temperature vortex dipole pair formation/breaking, and negative temperature same sign vortex cluster formation/breaking. A schematic summarising the vortex phases is shown in Fig.~\ref{Fig:F4}.
For two-dimensional neutral Coulomb gas with equal number of vortices and antivortices without core structure (true point vortices) \cite{salzberg1963equation, hauge19712}, the vortex dipole pair collapse transition occurs at a critical temperature $\beta_d  = 4\pi/\rho_s\Gamma^2 $, where $\rho_s$ is the fluid density, $\Gamma = h/m$ the circulation with $h$ and $m$ the Planck’s constant and atom mass, respectively. By accounting for the non-zero vortex core size, this transition temperature shifts \cite{viecelli1995equilibrium}, towards the Berezinskii--Kosterlitz--Thouless (BKT) critical temperature $\beta_{\rm{BKT}} =  2\beta_d$ \cite{berezinskii1971destruction, kosterlitz1973ordering, berezinskii1972destruction}. 

The negative temperature transition referred to as supercondensation or Einstein--Bose condensation (EBC) \cite{kraichnan1967inertial,kraichnan1980two,valani2018einstein}, corresponds to the condensation of like-sign Onsager vortex clusters. The critical temperature of this transition is $\beta_{\rm{EBC}} = -4\beta_d/N_v$ \cite{kraichnan1980two, viecelli1995equilibrium} for neutral vortex configurations and $\beta_{\rm{EBC}} = -2\beta_d/N_v$ for a system of single sign vortices. 

These temperatures serve as significant reference points characterizing the vortex configurations illustrated in Fig.~\ref{Fig:F4}. In between the two low entropy ordered phases the vortices are seemingly randomly distributed with their configurational entropy being maximized at $\beta = 0$. The top and bottom rows in Fig.~\ref{Fig:F4} show representative vortex configurations for the cases of $N_v = 10$ all clockwise (single sign) circulation and $N_v = 20$ with equal number of clockwise and anticlockwise (neutral) circulations, respectively. 

Our main goal in this paper is to apply machine learning technology to investigate if a simple unsupervised learning with neural networks is capable of correctly identifying the three significant vortex phases of matter summarised in Fig.~\ref{Fig:F4}. 

%%%%%%%%%%%%%%%%%%%%%%%%%%%%%%%%%%%%%%%%%%%%%%%%%%%%%%%%%%%%%%%%%%%
\subsection{Machine learning classification of vortex configurations} \label{Sec:ML}

Machine learning algorithms are used for categorization of data sets based on occurrences of common sets of features. Possible features include continuous, binary and categorical. Supervised machine learning requires additional knowledge (a training set) to supplement the data where as in an unsupervised machine learning approach the data set is provided for categorization as is without additional supporting information \cite{kotsiantis2007supervised}. 

The self-organising map (SOM) \cite{kohonen1982self,kohonen2007kohonen,kohonen2013essentials} is a case of unsupervised artificial neural networks (ANN) successfully applied in areas of data clustering, complex data visualization and for image processing. Here we apply a SOM algorithm to classify vortex configurations. For this purpose we employ a machine learning framework implemented in $\mathtt{MATLAB}$. In unsupervised learning, test images containing vortex positions are provided to the classifier and the self-organising maps cluster the data based on the detected similarity and topology \cite{kohonen1982self,kohonen2007kohonen,kohonen2013essentials}. First, the test images that are used for training the classifier need to be preprocessed. For this purpose, we use the bag of features (BoF) model to construct the feature vector (a histogram of discrete features detected in an image). The feature vectors are then used for training the SOM classifier model. These machine learning models used in this work are briefly described below.

%%%%%%%%%%%%%%%%%%%%%%%%%%%%%%%%%%%%%%%%%%%%%%%

\subsubsection{Bag of features model}\label{sec:MLbof}
Detecting robust image features forms the basis for accurate object recognition \cite{pena2011comparative}. The features of a digital image such as shapes, colour, texture, and the locations of these local features inside the image are properties that allow the image to be differentiated from other images in the database. Representing an image through its pixel values results in a very high dimensional matrix, which is not appropriate for image classification/recognition \cite{jahne1999handbook}. Therefore we instead extract the local features of digital images containing vortex positions to classify them in categories.

\begin{figure*}[]
   \centering
    \includegraphics[width=1.6\columnwidth]{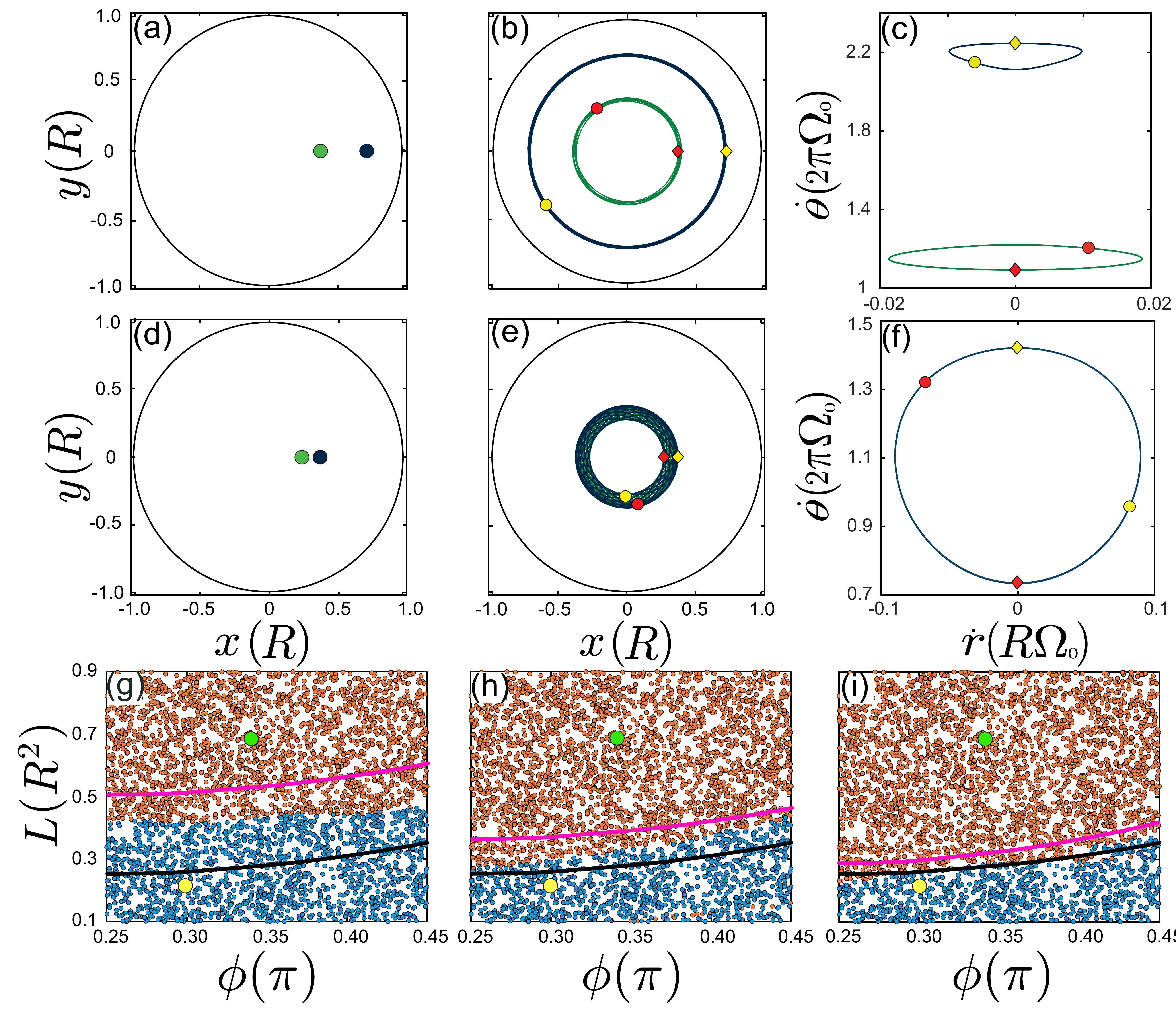}
    \caption{Unsupervised machine learning classification of vortex states for the case of two same sign vortices. Frames (a) and (d) show vortex configurations corresponding to the initial conditions 
    $\phi/\pi = 0.349$, $L = 0.69R^2$, $\theta_{21} = \pi$ and $\phi/\pi = 0.30$, $L = 0.22R^2$, $\theta_{21} = \pi$, respectively. Frames (b) and (e) show the vortex trajectories in real space corresponding to the initial configurations of (a) and (d) and integrated for the duration $60 \Omega_0^{-1}$. The frames (c) and (f) show the velocity space representations of the vortex dynamics in (b) and (e), respectively. In (b),(c),(e), and (f) the initial and final vortex positions are shown using lozenge and circular markers, respectively. The bottom row shows the machine learning classification of the data based on the vortex position data (first column), the vortex trajectory data (second column) and the vortex velocity data (third column). The green and yellow markers in (g), (h), and (i) label the two categories requested in the unsupervised machine learning classification and are obtained for a set of $4000$ initial conditions in $(\phi, L)$ space. The non overlapping vortex trajectories (b) are categorised as red markers and the overlapping trajectories (e) as blue markers. The black curves show the theory prediction for the phase boundary between the overlapping and non-overlapping vortex states \cite{murray2016hamiltonian}. The pink curves are the boundary generated using machine classification between the overlapping and non-overlapping vortex states with $2000$ initial configurations.}
    \label{Fig:F5}
\end{figure*}

The extracted features are influential when forming a signature, such as a bag of visual words (BoVW) depiction, for an image \cite{mansoori2013bag}. Bag of visual words is a simplified approach for extracting image content for machine learning classification \cite{bedo2016endowing}. It represents the images with orderless collections of local image features. Here we use the (BoF) model to construct the visual words for a given image. The concept of BoF is analogous to the \emph{bag of words (BoW)} \cite{suma2020novel, zhang2010understanding} used for representing a text document. The BoF model applies a similar methodology, but instead of \emph{words} it uses the image features for analysing the image. The key goal is to create a visual vocabulary known as codebook, where the most common and strong image features are coded as codewords or visual phrases. A visual vocabulary is then formed by clustering the extracted features from a set of images. Each feature cluster constitutes a visual word. The image representation as a BoF is a histogram generated by a simple image codeword/visual word occurrences analysis. This model treats each image as a visual word frequency histogram based on a vocabulary that measures the spatial characteristics of all images in the database. 

The complete bag of visual words process is accomplished by using an inbuilt \texttt{MATLAB} function, which operates in a step by step approach. Firstly, it extracts the strongest image features using speeded-up robust features (SURF) algorithm \cite{bay2008speeded,csurka2004visual}. Then specific patterns, shapes and edges are detected by SURF algorithm, and around each point of interest, the descriptor generates the description of the local neighbourhood pixels mainly by the intensity distribution of nearby pixels \cite{csurka2004visual, oyallon2015analysis}. The detailed description about SURF algorithm can be found in \cite{mistry2017comparison, zhu5surf}. The matching image pairs and hence strongest image features can be identified by comparing the descriptor across all images in image set \cite{oyallon2015analysis}. Furthermore, the SURF algorithm constructs the visual vocabulary by extracting $80{\%}$ of the strongest features and clusters them into visual words using a k-means clustering algorithm \cite{csurka2004visual, likas2003global}. The k-means clustering follows a heuristic approach to construct initial clustering by selecting random k-centroids from the data set in a two-dimensional similar feature space, which represents the SURF features as points \cite{kanungo2002efficient, shukla2014review}. For each data point the clustering algorithm calculates the distance from all centroids and then assign its membership to the nearest k-centroid iteratively. After each iteration, the recalculation of new k-centroid is done by averaging all data points that are assigned to the clusters and the process is repeated until convergence \cite{csurka2004visual}. After this process is complete and the final centroids become visual words comprising the visual vocabulary \cite{csurka2004visual}, for the given image set. These processed image features are used for training the (SOM) classifier model to reveal the possible classifications.

%%%%%%%%%%%%%%%%%%%%%%%%%%%%%%%%%%%%%%%%%%%%%%%%%%%%%%%%%%%%%%%%%%%
\subsubsection{Self-organising map algorithm}
A Kohonen self organising map (SOM) \cite{kohonen1982self} is a popular unsupervised artificial neural network which is used to group the similar patterns such as; feature vectors or data items together \cite{pacella2016use}. It projects a high-dimensional input data onto low-dimensional array of nodes (neurons) \cite{vesanto2000clustering}. This mapping retains the topological relationships between the data domains. Consequently, the image of the data space tends to manifest clustering of input information and their relationships on the map. This algorithm helps to understand high-dimensional input data by clustering similar data together and by reducing its effective dimensions. Initially a random weight is assigned for each neuron and is placed in the feature space containing the input vectors of the testing images. Then one of the input vectors is randomly selected. For each input vector, its Euclidean distance to every weight vector is calculated, and the neuron with the closest matching weight vector is moved towards the input vector in the feature space. 
Also the neighbouring neurons within a certain radius are dragged toward the input vector \cite{kohonen2013essentials, kohonen2007kohonen}. This ‘nearest’ neuron is called the best matching unit (BMU) or the winning neuron. The neurons' positions are updated in each iteration and the process is repeated for each input data and over all iterations. The magnitude of these displacements decrease with the distance from the BMU and as the iteration proceeds. After considering each neuron and all iterations, eventually the entire neural network tends to approximate the input vector distribution. Finally, the similar data clustered together in one area and the dissimilar one grouped in a separate area.

For analysing the classification of vortices we deployed the bag of feature function for extracting image features. We used a grid method for picking the key point location in the feature extraction mechanism and used a block width to specify the scale of the feature. We employed the grid method to optimize memory requirements and computational time \cite{chu2020grid}, while maintaining the accuracy of classifier. However, the grid size selection is a crucial step as a scattered grid that corresponds to a low number of image features can lead to loss of the key information, and on the other hand a dense grid that corresponds to excessively many features, becomes computationally demanding and may result in irrelevant information. For classifying two same sign vortices (data set of $4000$ images) we opt the GridStep =\;[8\; 8], BlockWidth =\;[32\; 64\; 96\; 128] and vocabulary size $=500$ in \texttt{MATLAB}. For analysing the classification of more vortices for polarised and neutral systems (larger data sets of $\approx 30,000$ images) we customize the bag of feature model in order to reduce the memory consumption while maintaining the desired accuracy.

Specifically, in the context of this work for larger data sets ($\approx 30,000$ images of size $291 \times 291$ pixels per image), the parameters in the deployed \texttt{bagOfFeatures} function call were: VocabularySize $=250$, StrongestFeatures $=0.8$, PointSelection = Grid, GridStep =\;[20\; 20], and BlockWidth =\;[32\; 64\; 96\; 128]. We ran the SOM clustering algorithm for $800$ training iterations and clustered the output typically into four different classes by setting the dimension of network (number of neurons) accordingly.

%%%%%%%%%%%%%%%%%%%%%%%%%%%%%%%%%%%%%%%%%%%%%%%%%%%%
%%%%%%%%%%%%%%%%%%%%%%%%%%%%%%%%%%%%%%%%%%%%%%%%
\section{Results}
To investigate the feasibility of using machine learning to classify point vortex configurations, we begin by considering the minimal system of two same sign vortices. Using the obtained results as encouragement we then move on to consider separately larger polarised and neutral vortex systems.

\subsection{Two same sign vortices}
Vortex dynamics of two same sign vortices in Bose--Einstein condensates has been observed experimentally \cite{Neely2010a,Navarro2013a}. In a system with a circular boundary each possible two-vortex configuration is either of overlapping or non-overlapping type \cite{Navarro2013a,murray2016hamiltonian}. Figure~\ref{Fig:F5}(a) and (d) show two such configurations. Visually, configurations ~\ref{Fig:F5}(a) and (d) look similar and their key difference is revealed by the dynamics shown in respective frames (b) and (e). In (b) the vortex paths never cross and the configuration (a) corresponds to a non-overlapping type. In (e) the vortex paths share a region of phase space and the configuration (d) corresponds to an overlapping type. The topological distinctness of these two types of vortex configurations becomes even clearer when considering the velocity space representation \cite{murray2016hamiltonian}, shown in (c) and (f).

To study the phase boundary between the overlapping and non-overlapping phases \cite{Navarro2013a,murray2016hamiltonian}, we generated $4000$ initial vortex configurations in the $(\phi, L)$ space and used the unsupervised machine learning approach to classify the vortex configurations. The images used for the classification in all cases were 221x221 pixels, to provide sufficient resolution for the machine learning algorithms (especially for feature extraction process) to operate effectively. We trained the clustering SOM algorithm on the feature vectors with the image features extracted using the default parameters of the bag of features function. 

Figure~\ref{Fig:F5}(g) shows the resulting unsupervised machine learning classification with red and blue markers corresponding to the two classified categories. The green and yellow markers correspond to the images (a) and (d), respectively. The black curve shows the correct location of the phase boundary \cite{Navarro2013a,murray2016hamiltonian}. Although the configurations (a) and (d) are classified correctly, the location of the phase boundary lies at higher value of $L$, when compared with its correct value, for all considered values of $\phi$. 

In the second test, we trained the SOM clustering algorithm with images showing the full vortex trajectories, such as (b) and (e) instead of the initial vortex configurations. The corresponding classification result shown in (h) is very similar to the case (g) with an improvement in the accuracy of the location of the phase boundary. In the third case (i) we have taken the pre-processing of the data even further and have trained the SOM clustering algorithm with images showing the velocity space representation of the vortex trajectories. The classification in this case is in excellent agreement with the correct phase boundary.

\begin{figure*}[!t]
    \centering
    \includegraphics[width=2\columnwidth]{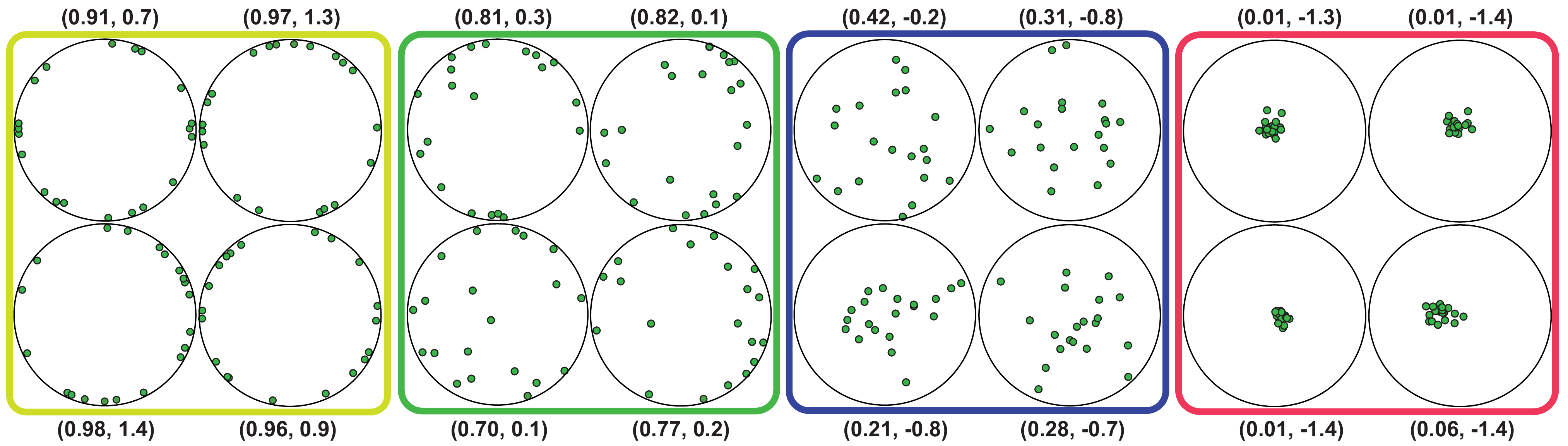}
\caption{Sixteen representative vortex configurations for the four color coded categories in Fig.~\ref{Fig:F7}. The $(L/N_v, \beta)$ values of each vortex configuration are shown closest to the respective images.}
\label{Fig:F6}
\end{figure*}
%%%%%%%%%%
\begin{figure}
   \centering
    \includegraphics[width=1\columnwidth]{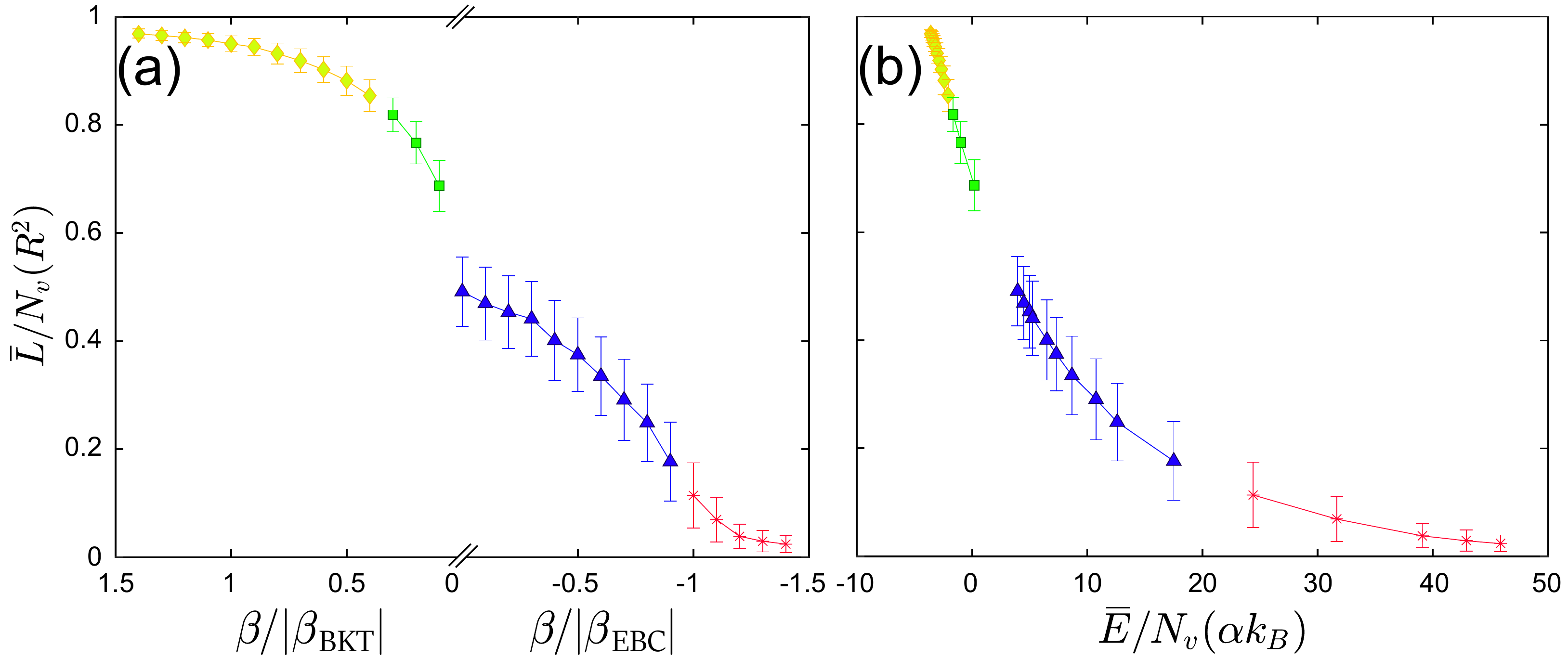}
    \caption{Unsupervised machine learning classification of point vortex configurations into four categories, corresponding to the different marker styles and colors. Frames (a) and (b) show, respectively, the vortex impulse per particle (${L}/N_v$),  as a function of inverse temperature ($\beta$) and the energy per vortex (${E}/N_v$). In (a) the x-axis is scaled by critical inverse temperature $|\beta_{\rm{EBC}}|$ and $|\beta_{\rm{BKT}}|$ for negative and positive temperatures, respectively. In both frames the y-axis is representing the average vortex impulse per particle ($\bar{L}/N_v$) and in (b) the x-axis is representing the average vortex energy per particle ($\bar{E}/N_v$). In both frames the classification is conducted for a system of $N_v = 20$ same sign vortices using an ensemble of $1000$ initial configurations. The error bars are one standard deviation statistical estimates.}
    \label{Fig:F7}
\end{figure}

The two main conclusions from these test cases are: (i) the unsupervised machine learning classification works for this problem remarkably well overall, and (ii) pre-processing the data before performing the classification can significantly improve the outcome of the classification. In addition, the number of sampled images affects the classification significantly, as demonstrated by the pink curves which show the identified boundary between the blue and red markers when $2000$ initial configurations are used. When the number of samples is increased to $4000$ the classified boundary shifts closer towards the correct (black curve) boundary, see Figs~\ref{Fig:F5} (g), (h) and (i). However, it is not always the case that the classification accuracy would continually improve with the increasing number of training samples since excessive training may lead to overlearning complications.

Although using the velocity space representation is ideal for this two-vortex problem, our numerics indicates that similar benefit over the vortex trajectory representation in the case of many vortex configurations is not realised. Furthermore, when the number of vortices in the system increases, the trajectory images such as (b) and (e) become increasingly over crowded and ultimately cannot be used for classification purposes as the whole image becomes covered densely for a fixed (resolution) trajectory line width. On the other hand, increasing the data resolution to resolve finer trajectory lines would rapidly lead to a memory bottleneck in computation. Consequently, for the remainder of this paper we are exclusively considering the plain vortex configurations such as (a) and (d), which also correspond to the most realistic experimental data in systems where real-time vortex tracking is not feasible.

%%%%%%%%%%%%%%%%%%%%%%%%%%%%%%%%%%%%%%%%%%%%
\subsection{Polarised vortex system}
%%%%%%%%%%%%%%%%%%%%%%%%%%%%%%%%%

The success of the machine learning approach in analysing the topological phase boundary of a two vortex system is encouraging and it motivates us to consider systems comprised of more vortices to assess the applicability of this machine model for classifying different phases of vortex matter. 

For this purpose, we generate point vortex configurations for $N_v = 20$ vortices of all same sign circulation using the Monte Carlo simulation method. We produce a sample of 1000 images per temperature point representing the statistically equivalent vortex configurations. A total of 29 temperature points uniformly distributed over the ranges $\beta =[1.4, 0]~\beta_{\rm BKT}$ and $\beta =[0, -1.4]~\beta_{\rm EBC}$, are considered.

Figure~\ref{Fig:F6} shows 16 example images of vortex configurations sampled at various temperatures. A collection of $29,000$ such images were sampled from the Monte Carlo run and were fed into the machine learning model. Then, we process the images as per the procedure explained in bag of feature model sec. \ref{sec:MLbof}. Thereafter, the processed image features are used to train the clustering algorithm.

%%%%%%%%%%
\begin{figure}[!t]
   \centering
    \includegraphics[width=\columnwidth]{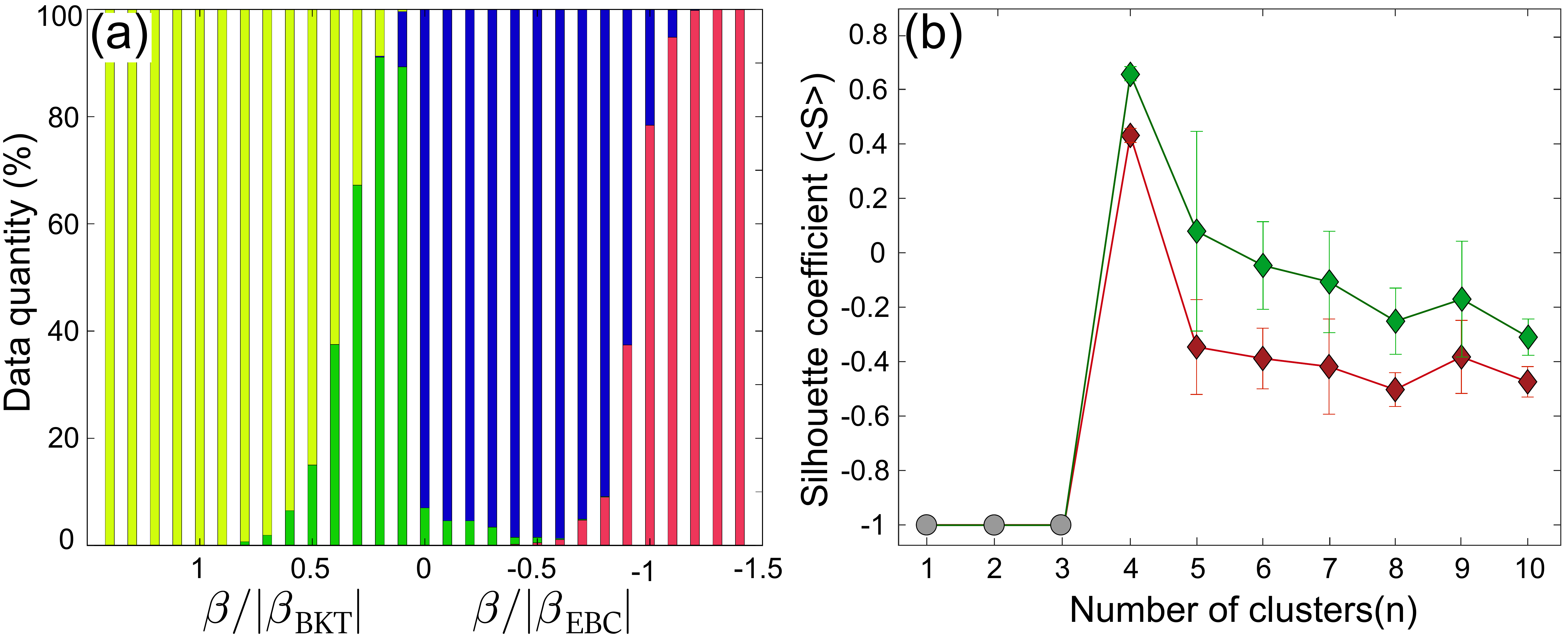}
    \caption{Frame (a) shows the percentage of images of the $N_v=20$  polarised vortex system that were classified by SOM to belong to each of the four color coded nodes as a function of the inverse temperature. Frame (b) shows the average silhouette value of the data set computed using k-means clustering as a function of number of clusters over $10$ different experiments. The red curve is the output of the same input data as in Fig~\ref{Fig:F8}(a) and green curve corresponds to silhouette analysis of 20 same sign vortices using an ensemble of 100 initial positions for 1 to 10 clusters. The silhouette analysis produce an overflow due to division by zero for cluster numbers $1-3$. We set those points to the minimum silhouette value of $-1$ as shown by grey circular markers.}
    \label{Fig:F8}
\end{figure}

The classification result with four vortex phases of polarised vortices is demonstrated using the average vortex impulse per particle ($\bar{L}/N_v$) as a function of inverse temperature ($\beta$) and the average energy per vortex ($\bar{E}/N_v$) as shown in Fig.~\ref{Fig:F7} (a) and (b), respectively. In (a) the x-axis is scaled by the critical inverse temperature $|\beta_{\rm{BKT}}|$ and $|\beta_{\rm{EBC}}|$ for positive and negative temperature, respectively. The error bars in (a) and (b) are one standard deviation of the statistical value of angular momentum per vortex, and the average energy per vortex, respectively. Samples of typical vortex configurations for the four color coded categories of Fig.~\ref{Fig:F7} are presented in Fig.~\ref{Fig:F6} and are encapsulated in corresponding colored frames. The $(L/N_v,\;\beta)$ values of each of the 16 vortex configurations are shown closest to the respective images.

\begin{figure*}[!t]
    \centering
    \includegraphics[width=2\columnwidth]{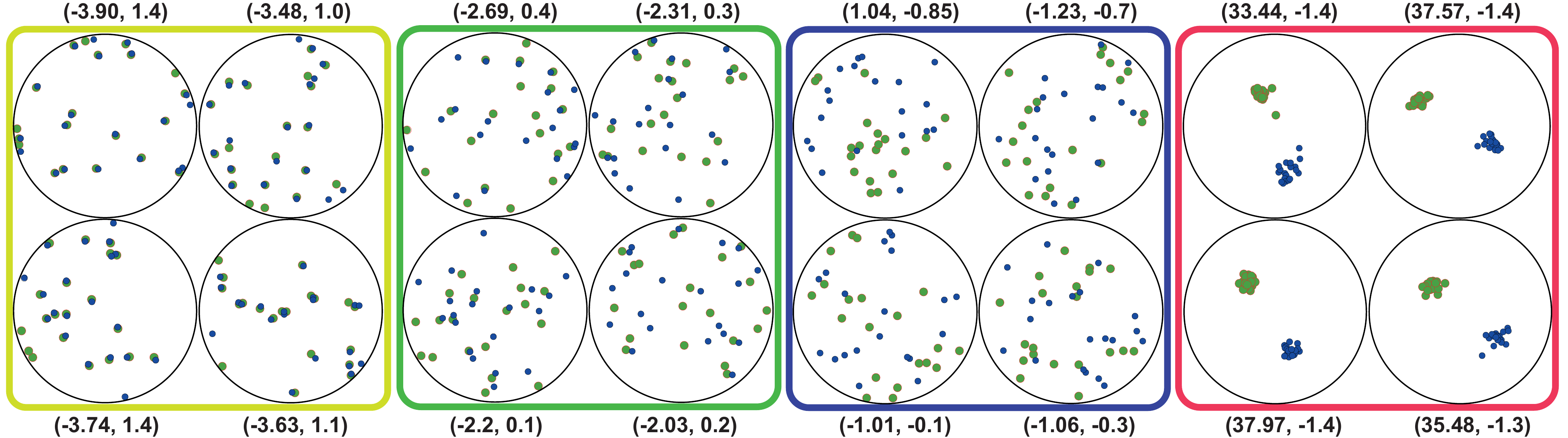}
\caption{Sixteen representative vortex configurations for the four color coded categories in Fig.~\ref{Fig:F10}. The $(L/N_v, \beta)$ values of each vortex configuration are shown closest to the respective images.}
\label{Fig:F9}
\end{figure*}

\begin{figure}[!ht]
   \centering
    \includegraphics[width=1\columnwidth]{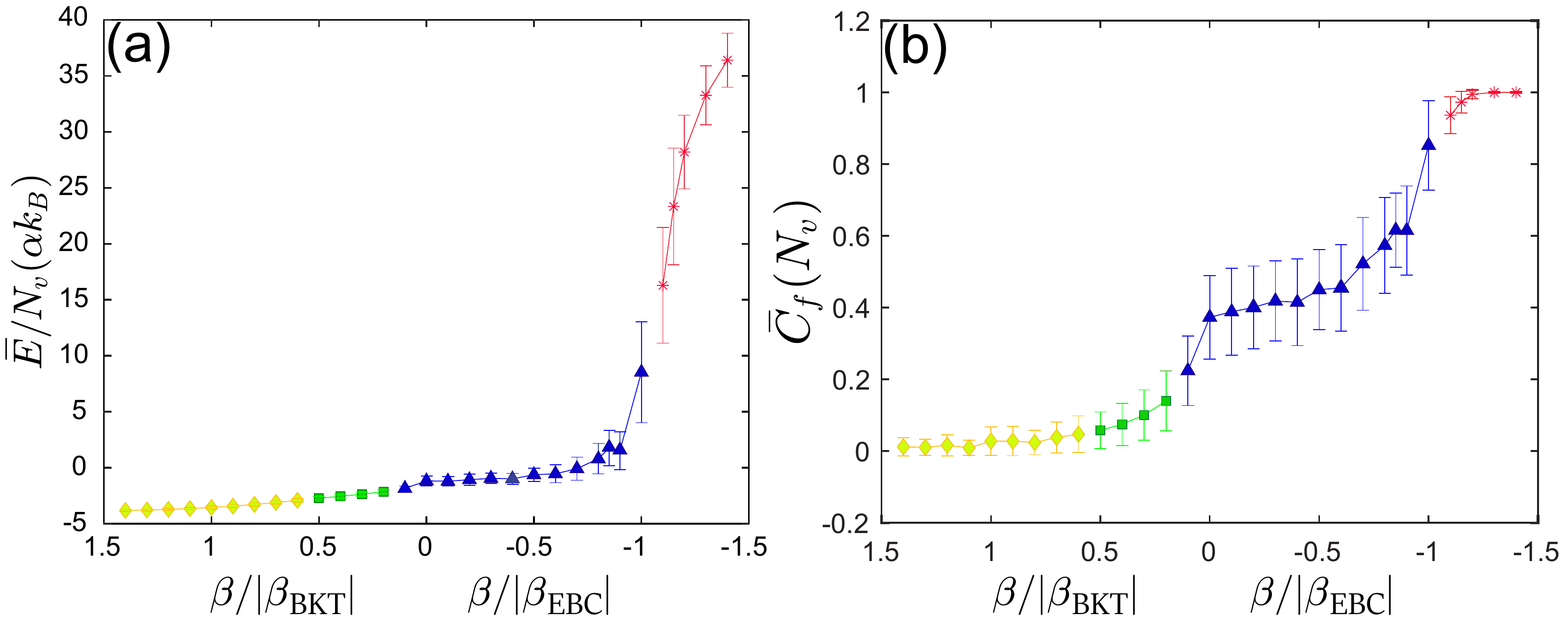}
    \caption{Unsupervised machine learning classification of point vortex configurations into four categories, corresponding to the different marker styles and colors. Frames (a) and (b) show, respectively, the energy per particle (${E}/N_v$) and vortex cluster fraction as a function of inverse temperature ($\beta$). In both frames the x-axis is scaled by critical inverse temperature $|\beta_{\rm{EBC}}|$ and $|\beta_{\rm{BKT}}|$ for negative and positive temperatures, respectively. In frame (a) the y-axis is representing the average energy per particle ($\bar{E}/N_v$) and in (b) the y-axis is representing the average cluster fraction ($\bar{C_f}$) in the units of number of vortices. In both cases the classification is conducted for a neutral vortex system of $N_v = 40$ vortices using an ensemble of $1000$ initial configurations. The error bars are one standard deviation statistical estimates.}
    \label{Fig:F10}
\end{figure}

Figure~\ref{Fig:F7} demonstrates that the unsupervised machine learning approach is able to distinguish four different vortex phases according to their vortex temperature. At high positive inverse temperature the classified yellow category corresponds to the vortex dipole phase, where the real vortices and their image vortices are paired across the boundary of the circular BEC as shown in the four configurations in the yellow box (Fig.~\ref{Fig:F6}). On increasing the temperature the vortex pairs begin to unbind from their images. For zero core point vortices this transition occurs at critical temperature $\beta/|\beta_{\rm{BKT}}| = 0.5$ and the machine learning classification seems to capture this transition. The green category corresponds to the high entropy, high positive temperature phase. When the temperature changes sign at $\beta = 0$ the output is classified in the blue category. Visually, many of the vortex configurations in the green and blue categories would be hard to distinguish where as the machine learning model has no difficulty in succeeding in this task. The fourth identified vortex matter phase is classified as a red category, which corresponds to the condensation of the well defined Onsager vortex clusters taking place at critical temperature $\beta/|\beta_{\rm{EBC}}| = -1$.

\begin{figure}[t]
   \centering
    \includegraphics[width=1\columnwidth]{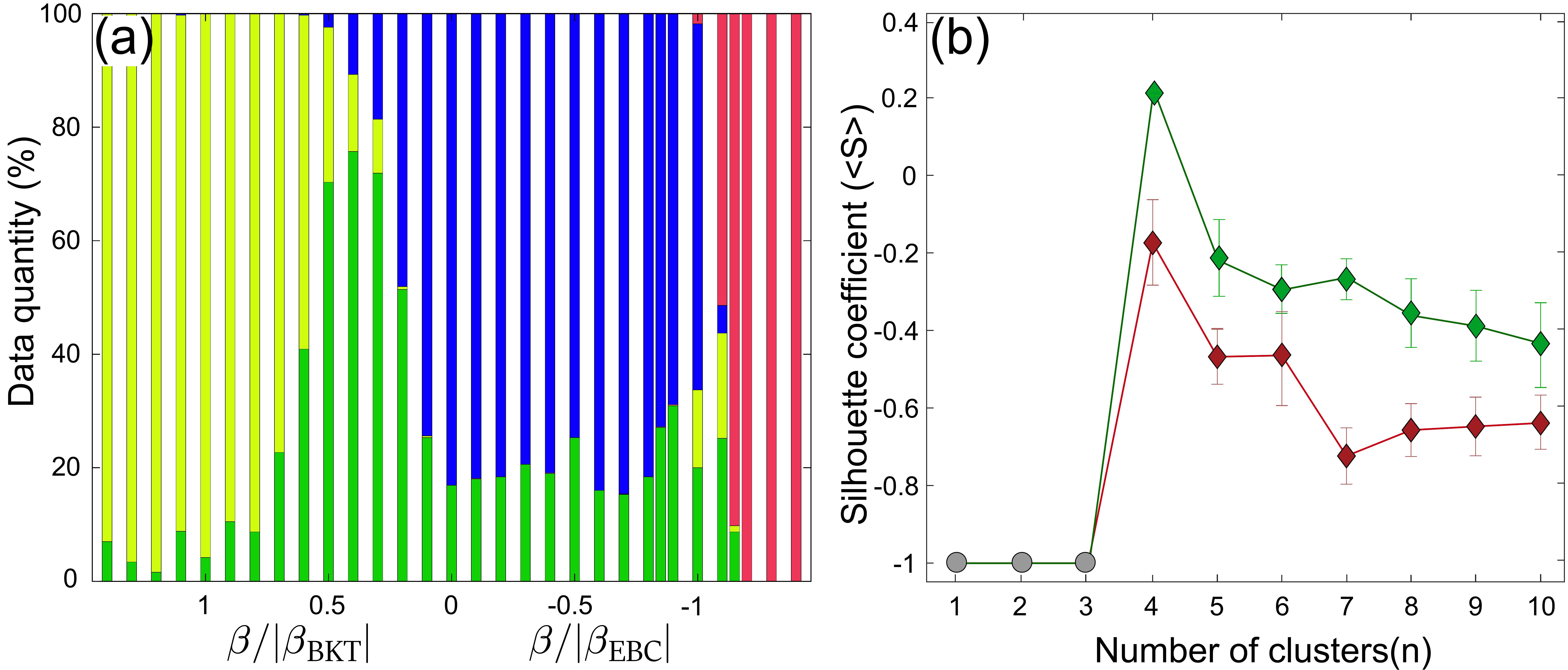}
    \caption{Frame (a) shows the fraction of images of neutral vortex system classified by SOM into the four nodes as a function of inverse temperature. The clustering was performed for a system of $N_v=40$ vortices using an ensemble of $1000$ initial configurations. Frame (b) shows the average of silhouette value for simulation results of k-means clustering as a function of number of clusters over $10$ experiments. The red curve is the output of the same input data as in Fig~\ref{Fig:F11}(a) and green curve corresponds to silhouette analysis of a system of $40$ neutral vortices using an ensemble of 100 initial positions. The silhouette analysis produce an overflow due to division by zero for cluster numbers $1-3$. We set those points to the minimum silhouette value of $-1$ as shown by grey circular markers.}
    \label{Fig:F11}
\end{figure}

To obtain deeper understanding of these classification results, Fig.~\ref{Fig:F8} (a) shows a histogram that counts the number of images classified into each of the four categories at each value of $\beta$. In accordance with the histogram, we represent each data point of Fig.~\ref{Fig:F7} using the majority color whose representation in the histogram exceeds $50\%$. To validate the SOM clustering output for the optimal number of four clusters for the tested data, we create a silhouette criterion clustering evaluation object using k-means clustering. The silhouette analysis provides the optimum cluster number (Silhouette Evaluation in $\mathtt{MATLAB}$) and evaluates the quality of clustering \cite{rousseeuw1987silhouettes}.

The silhouette value is a measure of how similar an item is to its corresponding cluster as compared to the separated clusters \cite{rousseeuw1987silhouettes}. The silhouette coefficient for $kth$ point is defined as $S_k = (b_k -a_k)/(\rm{max}(a_k, b_k))$, where $a_k$ is the average distance from the $kth$ point to all other points with in same cluster as $k$, and $b_k$ is the minimum average distance from the $kth$ object to objects in a different cluster. The silhouette coefficient $(S)$ for dataset is the mean silhouette  coefficients over the points. The value of $S$ ranges from $-1 \; \rm{to}\; +1$. A high value ($+1$) indicates that the sample is highly similar to its own cluster and quite distinct from other clusters. On the other hand, a low or negative $S$ value for many points indicates poor cluster compliance. That is, if many objects have a low silhouette value, then the clustering solution might have too many or too few clusters.

Fig.~\ref{Fig:F8}(b) presents the silhouette coefficient $(<S>)$  of clustering for $1$ to $10$ clusters averaged over $10$ experiments. The red curve is the average silhouette coefficient of clustering for the same data set that is clustered in Fig.~\ref{Fig:F8}(a) and green curve is the mean silhouette for the data clustering of a system of $20$ same sign vortices for $29$ temperature points and an ensemble of $100$ initial configurations at each temperature point. Thus, the total $2900$ images were created for this system and their features were extracted using custom extractor function for bagOfFeatures model in $\mathtt{MATLAB}$. This function uses the default SURF feature extraction over a uniform grid of point locations at many scales. The used default values for this feature extraction for grid steps and multifeature scales are $8$ and $\;[1.6\; 3.2\; 4.8\; 6.4]$, respectively. Then using k-means clustering for those extracted features of $2900$ images we evaluate the optimum cluster number by performing silhouette analysis. For both data sets we run the experiment $10$ times for 1 to 10 clusters and then averaged the silhouette value over total number of experiments for each cluster number. The error bars are one standard deviation statistical estimates of silhouette coefficients. 

In Fig.~\ref{Fig:F8}(b), the mean silhouette value is higher (green curve) for the smaller data set (preprocessed using custom extractor function with dense grid for BoF) than the larger data set (preprocessed with grid step size $[20\;20]$) but for both red and green curves the highest silhouette value occurs at four clusters, suggesting that the optimal number of clusters to be employed is four. This justifies our choice of four neurons in SOM classification, as using either more or fewer number for clusters leads to poorer clustering as a result of lower silhouette value. As such, the unsupervised machine learning classification is not only able to identify the boundaries of the four temperature regions but is also able to identify the correct number of physically meaningful regions.

%%%%%%%%%%%%%%%%%%%%%%%%%%%%%%%%%
\subsection{Neutral vortex system}

%%%%%%%%%%%%%%%%%
Having successfully classified the single sign vortices (polarised vortex fluid) into 4 temperature regions, we next repeat the analysis for the case of a neutral vortex system (with equal number of vortices and antivortices) having in total $N_v=40$ vortices. As for the polarised vortex systems, the test images of neutral vortex system using an ensemble of $1000$ initial configurations were produced using Monte Carlo simulation at $31$ temperature points. For machine learning classification the algorithm is trained on the feature vectors of images containing positions of vortices and antivortices plotted in green and blue circular markers as shown through $16$ representative images in Fig.~\ref{Fig:F9}. As in the case of polarized vortex system, here we also clustered the data into four categories. 

The quantitative classification results are again demonstrated using the average energy per particle ($\bar{E}/N_v$)$(\alpha k_B)$ and the average vortex cluster fraction ($\bar{C_f}(N_v)$) as a function of temperature ($\beta$) as shown in Fig.~\ref{Fig:F10} (a) and (b), respectively. The energy is in the units of $\alpha k_B$, where $\alpha = \rho_s \Gamma^2/4\pi k_B$. In both frames the x-axis is scaled by critical inverse temperature $|\beta_{\rm{BKT}}|$ and $|\beta_{\rm{EBC}}|$ for positive and negative temperature range, respectively.  
The error bars in (a) and (b) are one standard deviation in statistical value of energy per vortex and cluster fraction as a function of temperature, respectively. The vortex configurations for the four color coded categories in Fig.~\ref{Fig:F10} are presented in Fig.~\ref{Fig:F9} encapsulated by the corresponding colored boxes. The $(\beta,\; E/N_v)$ values of each vortex configuration are shown closest to the respective images. The vortices and antivortices are indicated by green (bigger) and blue (smaller) circular markers, respectively.

The unsupervised SOM algorithm again distinguishes excellently the four temperature regions. At high positive inverse temperature the yellow category corresponds to the lowest energy configurations within the pair collapse phase, where the vortices and antivortices are now paired up in the bulk (instead of forming edge states as in the polarised vortex fluid). A sample of corresponding configurations are shown in the yellow box of Fig.~\ref{Fig:F9}. On increasing the temperature the transition from vortex pair collapse to vortex unbinding takes place at $\beta/|\beta_{\rm{BKT}}| = 0.5$ and leads to the green category in Fig.~\ref{Fig:F10}. The negative temperature configurations around at $\beta = 0$ are classified into the blue category which correspond to the random negative absolute temperature vortex states. The red category is again identified as the Einstein--Bose condensate phase of Onsager vortices that emerges at  $\beta/|\beta_{\rm{EBC}}| = -1$.

The histogram of the classification is shown in Fig.~\ref{Fig:F11}(a) and as for the polarised vortex fluid, this was used for assigning the color coding in Fig.~\ref{Fig:F10} based on the $50\%$ criterion. Although the overall classification is again very good, configurations belonging to the green category are now observed infrequently at nearly all temperatures. The choice of four data clusters  is again justified by performing the silhouette analysis using k-means clustering for optimum cluster number evaluation. The resulting analysis of average silhouette value for 1 to 10 clusters over $10$ trials is shown in Fig.~\ref{Fig:F11}(b). The red curve corresponds to the average of mean silhouette value of clustering data (features extracted using $[20\;20]$ grid and $250$ visual words in bag of features function from $31,000$ input images) as used in Fig.~\ref{Fig:F11}(a). The green curve corresponds to the average silhouette value of clustering data ($3100$ input images with $100$ ensembles at each temperature point). For this smaller data set the features were extracted similarly using custom feature extractor function as explained for the data set of $2900$ images in the polarised vortex system. The result shows that the smaller data set whose features were extracted using a dense grid has higher silhouette value than the larger data (preprocessed with sparse grid), yet the highest silhouette coefficient occurs at four cluster in both cases. This justifies our choice of four neurons in SOM clustering as an optimal value for this data. In addition to this the low silhouette value (red curve) in Fig.~\ref{Fig:F11}(b) can be understood by inspecting the vortex configurations for larger data system. It is clear that small fraction of vortex dipoles are present at nearly all temperatures leading to the green category stretching over to the other temperature regions. To mitigate this issue, a further improvement of both training data and machine model, for example, high-quality vortex configuration images, additional thermalization points in the testing data preparation, or more precise feature extraction, could be performed before feeding them to the classifier model. However, even with the minimal amount of pre-processing, the classification results, as shown in Fig.~\ref{Fig:F10} are remarkably good. Figure~\ref{Fig:F11} (b) shows that the highest silhouette value occurs at four clusters, illustrating that the 4 neuron case is the optimal case also for the neutral vortex system.

%%%%%%%%%%%%%%%%%%%%%%%%%%%%%%%%%%%%%%%%%%%%%%%%%%%%%%%
\section{Conclusions}
The main objective of this paper was to test the feasibility of using a simple unsupervised machine learning approach to search for new vortex phases of matter and to identify the corresponding transition temperatures. We demonstrated the success of this approach using only the vortex positions as the input information for the machine model, even though the overall machine learning classification performance depends on a number of factors including prepossessing of the input data, extracted features, number of input sample images, number of requested clusters, and number of iterations in clustering algorithm. 

In the first part, we considered two same sign vortices and trained a neural network using three types of preprocessed input images comprising the vortex positions, vortex trajectories, and velocity space images. In each case the boundary separating overlapped and non-overlapped phase space regions \cite{murray2016hamiltonian}, was successfully detected and the sample size dependence of the location of the phase boundary result was demonstrated.

When considering larger numbers of vortices in both polarised and neutral vortex configurations, the unsupervised artificial neural network displayed consistent classification results when compared with  previously known results. Specifically, both the positive temperature Kosterlitz--Thouless transition in a two-dimensional Coulomb gas (point-vortex model) and the negative temperature Einstein--Bose condensation of Onsager vortices transitions were successfully identified. 

Furthermore, with the aid of silhouette analysis, the unsupervised machine learning model was also able to self-generate information on the optimal number of clusters to be employed for the classification and thereby the number of distinct temperature regions in the provided data set.

Considering that identification of the Kosterlitz--Thouless transition was found to be challenging even for complex convolutions neural network under a supervised approach \cite{beach2018machine}, that our simplified unsupervised approach was able to detect the vortex binding-unbinding transition in this system shows promise for further applications of this methodology. 

In light of the demonstrated performance, unsupervised machine learning has the potential to widen our understanding of topological phases of two-dimensional vortex matter, and may find applications in discovering exotic underlying vortex features both in theoretical models and in laboratory experiments, especially in the context of the research on two-dimensional quantum turbulence.

\begin{acknowledgements}
We are grateful to Andrew Groszek and Rodney Polkinghorne for helpful discussions. 
This research was funded by the Australian Government through the Australian Research Council (ARC) Future Fellowship FT180100020.
\end{acknowledgements}

%apsrev4-2.bst 2019-01-14 (MD) hand-edited version of apsrev4-1.bst
%Control: key (0)
%Control: author (72) initials jnrlst
%Control: editor formatted (1) identically to author
%Control: production of article title (-1) disabled
%Control: page (0) single
%Control: year (1) truncated
%Control: production of eprint (0) enabled
%

%\bibliographystyle{apsrev4-2}
%\bibliography{bibliography}

\end{document}